\newcount\mgnf\newcount\tipi\newcount\tipoformule
\newcount\aux\newcount\piepagina\newcount\xdata
\mgnf=0
\aux=1           
\tipoformule=1   
\piepagina=1     
\xdata=0         
\def\Di{}

\ifnum\mgnf=1 \aux=0 \tipoformule =1 \piepagina=1 \xdata=1\fi
\newcount\bibl
\ifnum\mgnf=0\bibl=0\else\bibl=1\fi
\bibl=0

%
%
%
%
\ifnum\bibl=0
\def\ref#1#2#3{[#1#2]\write8{#1@#2}}
\def\rif#1#2#3#4{\item{[#1#2]} #3}
\fi

\ifnum\bibl=1
\openout8=ref.b
\def\ref#1#2#3{[#3]\write8{#1@#2}}
\def\rif#1#2#3#4{}

\fi

\def\9#1{\ifnum\aux=1#1\else\relax\fi}
\ifnum\piepagina=0 \footline={\rlap{\hbox{\copy200}\
$\st[\number\pageno]$}\hss\tenrm \foglio\hss}\fi \ifnum\piepagina=1
\footline={\rlap{\hbox{\copy200}} \hss\tenrm \folio\hss}\fi
\ifnum\piepagina=2\footline{\hss\tenrm\folio\hss}\fi

\ifnum\mgnf=0 \magnification=\magstep0
\hsize=13.5truecm\vsize=22.5truecm \parindent=4.pt\fi
\ifnum\mgnf=1 \magnification=\magstep1
\hsize=16.0truecm\vsize=22.5truecm\baselineskip14pt\vglue5.0truecm
\overfullrule=0pt \parindent=4.pt\fi

\let\e=\varepsilon \let\z=\zeta 
\let\th=\vartheta\let\k=\kappa \let\l=\lambda \let\m=\mu \let\n=\nu
 \let\p=\pi  \let\s=\sigma \let\t=\tau
 \let\f=\varphi  
  \let\D=\Delta 
 \let\P=\Pi  
  
{\count255=\time\divide\count255 by 60 \xdef\oramin{\number\count255}
\multiply\count255 by-60\advance\count255 by\time
\xdef\oramin{\oramin:\ifnum\count255<10 0\fi\the\count255}}
\def\ora{\oramin }

\ifnum\xdata=0
\def\data{\number\day/\ifcase\month\or gennaio \or
febbraio \or marzo \or aprile \or maggio \or giugno \or luglio \or
agosto \or settembre \or ottobre \or novembre \or dicembre
\fi/\number\year;\ \ora}
\def\Di{\number\day\kern2mm\ifcase\month\or gennaio \or
febbraio \or marzo \or aprile \or maggio \or giugno \or luglio \or
agosto \or settembre \or ottobre \or novembre \or dicembre
\fi\kern0.1mm\number\year}
\else
\def\data{\Di}
\fi

\setbox200\hbox{$\scriptscriptstyle \data $}
\newcount\pgn \pgn=1
\def\foglio{\number\numsec:\number\pgn
\global\advance\pgn by 1} \def\foglioa{A\number\numsec:\number\pgn
\global\advance\pgn by 1}
\global\newcount\numsec\global\newcount\numfor \global\newcount\numfig
\gdef\profonditastruttura{\dp\strutbox}
\def\senondefinito#1{\expandafter\ifx\csname#1\endcsname\relax}
\def\SIA #1,#2,#3 {\senondefinito{#1#2} \expandafter\xdef\csname
#1#2\endcsname{#3} \else \write16{???? ma #1,#2 e' gia' stato definito
!!!!} \fi} \def\etichetta(#1){(\veroparagrafo.\veraformula) \SIA
e,#1,(\veroparagrafo.\veraformula) \global\advance\numfor by 1
\9{\write15{\string\FU (#1){\equ(#1)}}} \9{ \write16{ EQ \equ(#1) == #1
}}} \def \FU(#1)#2{\SIA fu,#1,#2 }
\def\etichettaa(#1){(A\veroparagrafo.\veraformula) \SIA
e,#1,(A\veroparagrafo.\veraformula) \global\advance\numfor by 1
\9{\write15{\string\FU (#1){\equ(#1)}}} \9{ \write16{ EQ \equ(#1) == #1
}}} \def\getichetta(#1){Fig.  \verafigura \SIA e,#1,{\verafigura}
\global\advance\numfig by 1 \9{\write15{\string\FU (#1){\equ(#1)}}} \9{
\write16{ Fig.  \equ(#1) ha simbolo #1 }}} \newdimen\gwidth \def\BOZZA{
\def\alato(##1){ {\vtop to \profonditastruttura{\baselineskip
\profonditastruttura\vss
\rlap{\kern-\hsize\kern-1.2truecm{$\scriptstyle##1$}}}}}
\def\galato(##1){ \gwidth=\hsize \divide\gwidth by 2 {\vtop to
\profonditastruttura{\baselineskip \profonditastruttura\vss
\rlap{\kern-\gwidth\kern-1.2truecm{$\scriptstyle##1$}}}}} }
\def\alato(#1){} \def\galato(#1){}
\def\veroparagrafo{\number\numsec}\def\veraformula{\number\numfor}
\def\verafigura{\number\numfig}
\def\geq(#1){\getichetta(#1)\galato(#1)}
\def\Eq(#1){\eqno{\etichetta(#1)\alato(#1)}}
\def\eq(#1){\etichetta(#1)\alato(#1)}
\def\Eqa(#1){\eqno{\etichettaa(#1)\alato(#1)}}
\def\eqa(#1){\etichettaa(#1)\alato(#1)}
\def\eqv(#1){\senondefinito{fu#1}$\clubsuit$#1\write16{No translation
for #1} \else\csname fu#1\endcsname\fi}
\def\equ(#1){\senondefinito{e#1}\eqv(#1)\else\csname e#1\endcsname\fi}
\openin13=#1.aux \ifeof13 \relax \else \input #1.aux \closein13\fi
\openin14=\jobname.aux \ifeof14 \relax \else \input \jobname.aux
\closein14 \fi \9{\openout15=\jobname.aux} \newskip\ttglue


\font\titolo=cmbx10 scaled \magstep1
\font\ottorm=cmr8\font\ottoi=cmmi7\font\ottosy=cmsy7
\font\ottobf=cmbx7\font\ottott=cmtt8\font\ottosl=cmsl8\font\ottoit=cmti7
\font\sixrm=cmr6\font\sixbf=cmbx7\font\sixi=cmmi7\font\sixsy=cmsy7

\font\fiverm=cmr5\font\fivesy=cmsy5\font\fivei=cmmi5\font\fivebf=cmbx5
\def\ottopunti{\def\rm{\fam0\ottorm}\textfont0=\ottorm%
\scriptfont0=\sixrm\scriptscriptfont0=\fiverm\textfont1=\ottoi%
\scriptfont1=\sixi\scriptscriptfont1=\fivei\textfont2=\ottosy%
\scriptfont2=\sixsy\scriptscriptfont2=\fivesy\textfont3=\tenex%
\scriptfont3=\tenex\scriptscriptfont3=\tenex\textfont\itfam=\ottoit%
\def\it{\fam\itfam\ottoit}\textfont\slfam=\ottosl%
\def\sl{\fam\slfam\ottosl}\textfont\ttfam=\ottott%
\def\tt{\fam\ttfam\ottott}\textfont\bffam=\ottobf%
\scriptfont\bffam=\sixbf\scriptscriptfont\bffam=\fivebf%
\def\bf{\fam\bffam\ottobf}\tt\ttglue=.5em plus.25em minus.15em%
\setbox\strutbox=\hbox{\vrule height7pt depth2pt width0pt}%
\normalbaselineskip=9pt\let\sc=\sixrm\normalbaselines\rm}

\catcode`@=11
\def\footnote#1{\edef\@sf{\spacefactor\the\spacefactor}#1\@sf
\insert\footins\bgroup\ottopunti\interlinepenalty100\let\par=\endgraf
\leftskip=0pt \rightskip=0pt \splittopskip=10pt plus 1pt minus 1pt
\floatingpenalty=20000
\smallskip\item{#1}\bgroup\strut\aftergroup\@foot\let\next}
\skip\footins=12pt plus 2pt minus 4pt\dimen\footins=30pc\catcode`@=12
\let\nota=\ottopunti

\newdimen\xshift \newdimen\xwidth \newdimen\yshift

\def\ins#1#2#3{\vbox to0pt{\kern-#2 \hbox{\kern#1
#3}\vss}\nointerlineskip}

\def\eqfig#1#2#3#4#5{ \par\xwidth=#1
\xshift=\hsize \advance\xshift by-\xwidth \divide\xshift by 2
\yshift=#2 \divide\yshift by 2 \line{\hglue\xshift \vbox to #2{\vfil #3
\includegraphics{#4.ps} }\hfill\raise\yshift\hbox{#5}}} 

\def\8{\write13}


\def\V#1{{\,\underline#1\,}}
\def\T#1{#1\kern-4pt\lower9pt\hbox{$\widetilde{}$}\kern4pt{}}
\let\dpr=\partial\def\Dpr{{\V\dpr}} \let\io=\infty\let\ig=\int
\def\fra#1#2{{#1\over#2}}\def\media#1{\langle{#1}\rangle}\let\0=\noindent
\def\guida{\leaders\hbox to 1em{\hss.\hss}\hfill}
\def\tende#1{\ \vtop{\ialign{##\crcr\rightarrowfill\crcr
\noalign{\kern-1pt\nointerlineskip} \hglue3.pt${\scriptstyle%
#1}$\hglue3.pt\crcr}}\,} \def\otto{\
{\kern-1.truept\leftarrow\kern-5.truept\to\kern-1.truept}\ }

\def\pagina{\vfill\eject}

\def\st{\scriptscriptstyle}
\def\*{\vskip0.3truecm}

\def\lis#1{{\overline #1}}\def\eg{\hbox{\it e.g.\ }}
\def\ap{\hbox{\it a priori\ }}
\def\ie{\hbox{\it i.e.\ }}

\def\fiat{{}}
\def\\{\hfill\break} \def\={{ \; \equiv \; }}

\def\annota#1{\footnote{${{}^{\bf#1}}$}}
\ifnum\aux=1\BOZZA\else\relax\fi
\ifnum\tipoformule=1\let\Eq=\eqno\def\eq{}\let\Eqa=\eqno\def\eqa{}
\def\equ{{}}\fi
\def\defi{\,{\buildrel def \over =}\,}

\def\1{\ifnum\mgnf=0\pagina\else\relax\fi}
\def\W#1{#1_{\kern-3pt\lower6.6truept\hbox to 1.1truemm
{$\widetilde{}$\hfill}}\kern0pt}

\def\LL{{\cal L}}

\def\EE{{\cal E}}\def\xx{{\V x}}
\def\cfr{{\it c.f.r.\ }}
\def\kk{{\V k}}\def\uu{{\V u}}
\def\VV#1{{\underline
#1}_{\kern-3pt$\lower7pt\hbox{$\widetilde{}$}}\kern3pt\,}

\def\FINE{
\*
\0{\it Internet:
Authors' preprints downloadable (latest version) at:

\centerline{\tt http://ipparco.roma1.infn.it}
\centerline{(link) \tt http://www.math.rutgers.edu/$\sim$giovanni}

\*
\sl e-mail: giovanni.gallavotti@roma1.infn.it
}}
\fiat

\def\AA{{\cal A}}\def\CS{{\cal S}}\def\KK{{\cal K}}

\fiat

\def\CH{chaotic hypothesis\ }

\centerline{\titolo Fluctuations and entropy driven space--time
}
\centerline{\titolo intermittency in Navier--Stokes fluids.}
\*\*
\centerline{\it Giovanni Gallavotti}
\*
\centerline{Fisica, Universit\`a di Roma 1}
\centerline{P.le Moro 2, 00185 Roma, Italia}
\*\*
\0{\bf Abstract: \it We analyze the physical meaning of fluctuations
of the phase space contraction rate, that we also call entropy
creation rate, and its observability in space--time intermittency
phenomena. For concreteness we consider a Navier--Stokes fluid.}
\*\*

\0{\bf\S1. The chaotic hypothesis in turbulence.}
\numsec=1\numfor=1\*

Consider a Navier--Stokes (NS) fluid in a container $V$ which we
take, for simplicity, cubic with periodic boundary conditions and
subject to a constant volume force $f \,\V\f(\xx)$ with $\max
|\V\f(\xx)|=1$ and with only Fourier harmonics corresponding to large
wavelength of the order of the linear size $L$ of $V$.

The viscosity will be denoted $\n$, but it is convenient to rescale
space, time, velocity and pressure to write the equations in
dimensionless form as

$$\dot{{\uu}}+ R\,\W u\cdot\W\dpr\,\uu=\D\,\uu-\Dpr p+\V\f, \qquad
\Dpr\cdot\uu=\V0,\qquad R=f L^3\n^{-2}\Eq(1.1)$$
in a container of side $L=1$, where $R$ is the {\it Reynolds
number}. We can suppose that $\ig_V\V u\,d\xx=\V0$ (because of
translation invariance).

We assume the \CH
\*

\0{\bf Chaotic hypothesis: \it Asymptotic motions of a turbulent flow
develop on an attracting set $\AA$ in phase space on which time
evolution $\V u\to S_t\,\uu$ can be regarded as a transitive Anosov
system for the purposes of computing time averages in stationary
states.}
\*

Here we investigate some assumptions under which the hypothesis
acquires some non trivial predictive value with implications that can
have experimental relevance. For earlier reviews on the chaotic
hypothesis see [Ga98a], [Ga96d], [Ga99a]. A recent one is [Ru99a].

\*
\0{\bf\S2. The OK41 cut--off.}
\numsec=2\numfor=1\*

Anxiety often mars the beginning of any discussion on the NS
equations: it is a fact that to date there is {\it no theory} that
allows a constructive solution of the equations via a controlled
approximation scheme. Nevertheless most people rapidly recover and
adopt the viewpoint that ``physically there is an effective
ultraviolet cut--off'' and the NS equations can be reduced to ordinary
equations:
\*

\0{\bf The OK41 cut--off hypothesis: \it There exists $\k_0>0$ such
that if the NS equation, \equ(1.1), is truncated in momentum space at
$K(R)=R^{\k_0}$ (or higher) then the physically relevant predictions
are not affected.}
\*

The OK41 theory, see [LL71], assigns to $\k_0$ the value
$3/4$. Therefore the flows of physical interest should be described by
\equ(1.1) truncated at $|\kk|\le K(R)$, \ie

$$\dot{{\V u}}_\kk+i\,R\,\sum_{\kk_1+\kk_2=\kk\atop |\kk_j|\le K(R)}
\W u_{\kk_1}\cdot\W k\,\P_\kk \V u_{\kk_2}\,=\,-\kk^2\V u_\kk+\V\f_\kk
\Eq(2.1)$$
where $\V u(\xx)\defi\sum_{\kk\ne\V0} e^{i\,\kk\cdot\xx}\uu_{\kk}$ and
$\P_\kk$ is the projection orthogonal to $\kk$; $\kk=2\p\V n$ with $\V
n$ an integer components vector.

Equation \equ(2.1) admits an \ap bound on the energy $\dot
E/2=-\sum_{\kk}\kk^2|\V u_\kk|^2+\sum_\kk \lis\f_\kk\cdot \V u_\kk$
which implies that asymptotically in time the energy is bounded by
$E\le 2||\V\f||^2/(2\p)^4$.

We shall call $\m_R$ the ``{\it statistics}'' of the NS equation defined
as the probability distribution on {\it phase space} (\ie on the space
of the velocity fields $\{\V u_\kk\},\,|\kk|<K(R)$) which describes, at
Reynolds number $R$, the stationary state averages of observables
$F(\uu)$, \ie the probability distribution such that

$$\lim_{T\to\io} T^{-1}\ig_0^T F(S_t\uu) dt=\ig F(\V v)\m_R(d\V
v)\Eq(2.2)$$
for almost all initial data $\uu$. The distribution $\m_R$ exists and
is unique because of the \CH and it is also called the SRB distribution
of the stationary state of \equ(2.2).
\*

\0{\bf3.  NS and GNS equations: viscosity and vorticity ensembles.
Equivalence.}
\numsec=3\numfor=1\*

For the purposes of a conceptual analysis stressing the analogy with
the theory of ensembles in statistical mechanics we temporarily
introduce a {\it control parameter} $\l$ in front of the Laplacian in
\equ(1.1) and in front of the $-\kk^2\,\uu_\kk$ term in \equ(2.1):
bearing in mind, however, that we are interested in $\l=1$ {\it
only}. The stationary distributions will then depend on $\l,R$ and
will be denoted $\m_{\l,R}$ so that the previously introduced SRB
distribution $\m_R$, \equ(1.1), is $\m_R\=\m_{1,R}$ with the new
notations.
\*

The collection $\EE_{NS}$ of all the probability distributions $\m_{\l,R}$
is the collection of all the stationary states of the fluid, at
varying Reynolds number $R$. It is an {\it ensemble} in the sense of
statistical mechanics and it will be called the ``{\it viscosity
ensemble}'' for the NS equations.

The second idea is that the same fluid can be studied, rather than by
the NS equations \equ(2.1), by considering the Euler equations subject
to a dissipation mechanism that keeps the {\it vorticity}
$\CS\defi\ig_V (\W\dpr\uu)^2\,d\xx$ bounded. This ``thermostatting''
effect can be achieved by imposing various types of forces $\V{{th}}$ on
the system so that $\CS=\,const$ leading to

$$\dot{{\uu}}+ R\,\W u\cdot\W\dpr\,\uu=-\Dpr p+\V\f+\V{{th}}, \qquad
\Dpr\cdot\uu=\V0,\qquad R=f L^3\n^{-2}\Eq(3.1)$$
As an example we can consider the force obtained by imposing the
constraint that $\CS$ remains identically constant via Gauss'
principle of least effort, \cfr [Ga96b], appendix. It corresponds to,
\cfr [Ga96b]
$$\dot{{\uu}}+ R\,\W u\cdot\W\dpr\,\uu=-\Dpr p+\V\f+\n_G(\uu)\,\D\uu,
\qquad
\Dpr\cdot\uu=\V0\Eq(3.2)$$
where $\n_G(\uu)$ is an easily determined multiplier defined so that
$\CS$ is {\it exactly} conserved, namely
$$\n_G(\uu)=\fra{\ig_V\big(\V\f\cdot\D\uu-R\D\uu\cdot(\W
u\cdot\W\dpr\uu)\big)\,d\xx} {\ig_V(\D\uu)^2\,d\xx}\Eq(3.3)$$
where $V$ is the container region.

The collection $\tilde \EE$ of the statistics $\tilde \m_{\CS,R}$ for
\equ(3.1) will be called the ``{\it vorticity ensemble}''. We establish a
correspondence between elements of the ensembles $\EE$ and $\tilde\EE$
by saying that two elements $\m_{\l,R}\in \EE$ and
$\m_{\CS,R}\in\tilde\EE$ are correspondent if
$$\tilde \m_{\CS,R}(\n_G)=\l\Eq(3.4)$$

We shall call ``{\it local}'' an observable $F(\uu)$ that depends only
on finitely many Fourier components of the velocity field $\uu$ (this
is {\it locality in momentum space}) and $\LL$ is the family of the local
observables. The analogy with statistical mechanics is quite manifest
if the following conjecture holds, [Ga95b],
\*

\0{\bf Equivalence conjecture: \it
Let $\m_{\l,R}\in \EE$ and $\m_{\CS,R}\in\tilde\EE$ be corresponding
elements of the viscosity and vorticity ensembles, \ie if $\CS$ and
$\l$ are related by \equ(3.4), then it is

$$\lim_{R\to\io} \fra{\tilde\m_{\CS,R}(F)}{\m_{\l,R}(F)}=1\Eq(3.5)$$
for all local observables $F\in\LL$ with non zero average.}
\*

In other words the statistics of the {\it irreversible} NS equation
\equ(1.1) and of the {\it reversible} ``Gaussian NS equation'' \equ(3.1),
called GNS equation, form two {\it equivalent ensembles of stationary
states} of the fluid. By ``reversible'' we mean that there is an
involutory map $I$ of phase space which {\it anticommutes with time
evolution}, \ie

$$I^2=1,\qquad I\,S_t= S_{-t} I\quad \hbox{ for all}\quad t\Eq(3.6)$$
and the GNS equations are reversible because $\n_G(\uu)$ is odd under
the transformation $I\uu(\xx)=-\uu(\xx)$. A similar conjecture has
been proposed in certain models of non equilibrium statistical
mechanics, [Ga96b], [Ru99b], [Ga99d].
\*

For reference purposes we write explicitly the GNS equations with the
OK41 cut--off

$$\dot{{\V u}}_\kk+i\,R\,\sum_{\kk_1+\kk_2=\kk\atop |\kk_j|\le K(R)}
\W u_{\kk_1}\cdot\W k\,\P_\kk \V u_{\kk_2}\,=\,-\n_G(\uu)\,
\kk^2\V u_\kk+\V\f_\kk
\Eq(3.7)$$
with the same notations of \equ(2.1).
\*

The analogy with equilibrium theory of ensembles is: the parameter $R$
plays the role of the {\it volume} while $\l$ that of the {\it
temperature} and $\CS$ that of the {\it energy}. Therefore the viscosity
ensemble is the analogue of the {\it canonical ensemble} and the vorticity
ensemble is the analogue of the {\it microcanonical ensemble}. The
$R\to\io$ is analogous to the ``{\it thermodynamic limit}'', [Ga99a].

We see also why it is useful to introduce the parameter $\l$: if we
stick to $\l=1$ then effectively we consider only a {\it single}
stationary state $\m_{1,\l}=\m_R$ and {\it not} an ensemble: this
state is ``the same'' (in the sense \equ(3.5)) as the state
$\tilde\m_{\CS_R,R}$ if $\CS_R$ is so defined that
$\tilde\m_{\CS_R,R}(\n_G)=1$. The parameter $\l$ will be set
to its physical value $1$ from now on.
\*

The conjecture of equivalence was proposed in [Ga97a] and discussed in
several other papers, see for instance [Ga97b]. It has been
investigated by simulations in [RS99] with results that seem moderately
satisfactory.
\*

\0{\bf\S4. Time reversal and fluctuation theorem.}
\numsec=4\numfor=1\*

We now consider the NS equation \equ(3.7) and try to find some of its
properties under the assumption that it is equivalent to the
corresponding GNS equation, \ie \equ(3.7) with $-\kk^2\uu_\kk$
replaced by $-\n_G(\uu)\kk^2\uu_\kk$. We assume the \CH and the OK41
cut--off and furthermore that
\*

\0{\bf Transitivity and axiom C: \it
Either the full ellipsoid in phase space

$$\{\uu\,|\,\sum_{|\kk|<R^{\k_0}}\kk^2|\uu_\kk|^2=\CS_R\}\Eq(4.1)$$
is densely visited by the evolutions of all data starting on it apart
from a zero volume set, or alternatively the evolution on this
ellipsoid verifies a geometric property called ``{\it axiom C}''.}
\*

Axiom C says that if the system is not transitive because there is an
attracting set $\AA$ that is {\it smaller} than the full phase space
(\ie the ellipsoid \equ(4.1) in this case) then, considering the simple
case in which this happens because in phase space there are just a non
dense attracting set and a repelling set (also not dense),
\*

\0(1) the attracting and the repelling sets are smooth manifolds and all
    their points, but a set of zero surface area, generate dense
    trajectories on them, and
\\
(2) the stable manifold of the points on the attracting set crosses
    transversally the repelling set and viceversa the unstable manifold
    of a point on the repelling set crosses transversally the
    attracting set.
\*

\0for details, which will not be really necessary here, we refer to
[BG98]. This implies that either the system is transitive or that its
restriction to the attracting set is transitive.

The interest of the Axiom C notion is that it is a geometric property
that has a remarkable consequence for systems admitting a time
reversal symmetry $I$ but with an attracting set $\AA$ {\it which is
not} the full phase space and, therefore, is mapped by $I$ onto a
repelling set $I\,\AA$ different from $\AA$. If the axiom holds one
can define, [BG98], a map $P:\,\AA\otto I\,\AA$, of the attracting set
$\AA$ on the repelling set $IA$ which commutes with time evolution and
with $I$ and

$$I\,P\, S_t=S_{-t}\, I\,P\Eq(4.2)$$
\ie the restriction of the transitive evolution $S_t$
to the attracting set $\AA$ {\it is still reversible}, although it is
such for a {\it new time reversal operation}, namely $I\,P$, see
[BG98], [Ga98b].
\*

{\it If a reversible evolution verifies axiom C and depends on a
parameter and, as the parameter varies, it develops an attracting set
$\AA\ne I\,\AA$ that is not the full phase space then the restriction
of the evolution to the attracting set is time reversible with respect
to a new time reversal symmetry.} 

In other words in axiom C systems time reversal symmetry $I$ {\it
cannot be really broken}: if there is a spontaneous breakdown (such
has to be considered the ``breaking'', as a parameter varies, of phase
space into an attracting set $\AA$ smaller than phase space and a
repelling set $I \,\AA$ different from $I\,\AA$, [Ga98b]) a new time
reversal $PI$ is ``spawned''.
\*

The axiom C property is stable under perturbations: changing slightly
parameters a system keeps this property if it has it to begin with.
The transitivity (or axiom C) property is relevant because of the
following theorem
\*

\0{\bf Theorem \it (fluctuation theorem): Let $-\s(\uu)$ be the
divergence of the GNS equations \equ(3.7) and let $-\media{\s}_+$ be
its stationary average with respect to the SRB distribution.
\\
Then the (dimensionless) quantity

$$p=\t^{-1}\ig_{-\t/2}^{\t/2} \fra{\s( S_t
\uu)}{\media{\s}_+}\,dt \Eq(4.3)$$
which we call ``{\sl average over a time span $\t$ of the
(dimensionless) phase space contraction at $\uu$}'' has a probability
of being in the interval $[p,p+dp]$ of the form
$\p_\t(p)\,dp=\,const\, e^{\z(p)\t+O(1)}$ with

$$\z(-p)=\z(p)-\media{\s}_+\,p\Eq(4.4)$$
for all $p$.}
\*

This theorem can be found in [GC95] for evolution maps and in [Ge98]
for flows (which is the version we use here): see also [Ru99a]. The
quantity $p$ depends on $\uu$.  The quantity $\media{\s}_+$ is also
called ``average entropy creation rate'' and $p=p(\uu)$ is the
dimensionless entropy creation rate averaged over a time $\t$ and in
the point $\uu$: see [An82], [Ru96], [GR97], [Ru99a]. We recall that
entropy in systems out of equilibrium is not defined (yet) so that
this name needs not be taken too seriously and it might eventually
reveal itself inappropriate. 

The above result should not be confused (as it is conceptually and
technically different) with other apparently similar statements, see
[CG99]. It was discovered as an experimental relation in a numerical
simulation, [ECM93], where the role of the SRB distributions and of time
reversal were also suggested to be a possible reason for its validity:
this led to its proof for Anosov maps in [GC95] and for flows in [Ge97].
\*

A key feature of the theorem is that it contains {\it no free
parameters}: its generality makes it a mechanical identity in the same
sense, although of course of not comparable importance, as the heat
theorem of Boltzmann, [Bo66], [Bo84], see also [Ga99a].

In the case of axiom C systems \equ(4.4) still holds, because the
evolution restricted to $\AA$ is transitive and reversible by
\equ(4.2), {\it but $\s$ has to be replaced by the contraction rate
$\s_0$ of the surface area of the attracting set $\AA$}. The
quantities $\s$ and $\s_0$ seem unrelated; however there are important
cases in which the total phase space contraction $\s$ and the
contraction of the surface element of the attracting set $\s_0$ are
proportional: $\s_0=\th_0\s$ with $\th_0$ a constant factor (or
varying on a slower time scale than $\s$ itself). Then
$\media{\s_0}_+=\th \media{\s}_+$ with $\th=\media{\th_0}_+$ and
\equ(4.4) becomes
$$\z(-p)=\z(p)-\th\,\media{\s}_+\Eq(4.5)$$
for all $p$. The GNS case is not among the (important) cases in which
$\s$ and $\s_0$ are proportional, see [DM96], [WL98],[BGG97] and [BG97],
although heuristic arguments can be given, [Ga97a], suggesting that
nevertheless a relation like \equ(4.5) might hold.

In some cases in which proportionality between $\s$ and $\s_0$ can be
established, at least on heuristic grounds, the proportionality factor
is just $1-d(\AA)/d$ if $d$ is the dimension of phase space and $d(\AA)$
is the dimension of the attractor, but unfortunately such cases are very
rare, [BGG97], [BG97].
\*

Finally it is worth writing explicitly the expression of the phase
space contraction $\s(\uu)$ for the equation \equ(3.7)
$$
\eqalign{
&\s(\uu)=(\sum_{|\kk|< K(R)}\kk^2)\, \n_G(\uu)-\big(\ig_V
\D\V\f\cdot\D\uu\,d\xx\big)\,\big(\ig_V [(\D\uu)^2-\cr
&- R\, \D\uu\cdot(\D(\W u\cdot\W\dpr \uu))-R (\D \W
u)\cdot(\D\uu)\cdot (\W\dpr \uu) -R\D\uu\cdot (\D\W\dpr \uu)\W u+\cr
&+\n(\uu)\D\uu\cdot \D^2\uu ]d\xx\,\big)\ / \ig_V
(\D\uu)^2\,d\xx\cr}\Eq(4.6)$$
In this expression (straightforwardly derived by imposing that $\CS$ is
exactly constant on motions verifying \equ(3.7)) the first term seems to
be the dominant one at large $R$ so that
$$\s(\uu)\simeq (\sum_{|\kk|< K(R)}\kk^2)\, \n(\uu)\defi
\KK(R)\,\n_G(\uu), \qquad \KK(R)\propto R^{3\k_0+2}\Eq(4.7)$$
which, if the side $L_0$ of the box is not $L=1$ would be written with
$\KK(R)$ replaced by
$\KK_{L_0}(R)=\sum_{|\kk|< K(R)}\kk^2$ with $\kk=2\p\V n/L_0$.
\*
\1

\0{\bf\S5. Fluctuation patterns and an extension of
Onsager--Machlup fluctuations theory.}
\numsec=5\numfor=1\*

A physical interpretation of the fluctuation theorem, when it holds,
can be found along with proposals for its test in experiments. We need
first some consequences of the (technique of proof of the) fluctuation
theorem.

Given an observable $H(\uu)$ we say that in its evolution observed
over a time interval of size $\t$ it {\it follows a pattern} $t\to
h(t)$ if $F(S_t\uu)=h(t)$ for $t\in[-\t/2,\t/2]$. We assume that $F$
has well defined time reversal parity $\e=\pm1$: $F(I\,\uu)=\e\,
F(\uu)$, for simplicity; and we say that the pattern $Ih(t)=\e\,h(-t)$
is the {\it time reversed pattern} of $h$.
\*

Fluctuation patterns are the main object of analysis in the theory of
Onsager--Machlup which deals with the probability of observing a
fluctuation pattern $h$ for an observable in the linear response
regime (\ie strictly speaking it deals with derivatives of various
quantities with respect to the strength of the forcing terms evaluated
at $0$ forcing).

The following theorem can be regarded a result of the same type {\it
without the restriction} that the system is in the linear response
regime.
\*

\0{\bf Theorem \it(entropy creation as a fluctuations driver):
\\
Consider a time reversible evolution verifying the chaotic hypothesis
and transitivity.  Let $H,K$ be two local observables (of given time
reversal parity) and denote $\m_{R,\t,p}$ the SRB distribution {\sl
conditioned to a (dimensionless) phase space average contraction $p$
over a time span $\t$}. Let $h,k$ be two fluctuation patterns for $H,K$
and let $Ih,Ik$ be their time reversal patterns. Then if
$\m_{R,\t,p}(\hbox{\rm pattern of $H$}=h)$ denotes the probability that
$H$ follows the pattern $h$ in the time span $\t$ in which the average
dimensionless phase space contraction is $p$ it is
$$\fra{\m_{R,\t,p}(\hbox{\rm pattern of $H$}=h)}{\m_{R,\t,p}(\hbox{\rm pattern
of K}=k)}=\fra{\m_{R,\t,-p}(\hbox{\rm pattern of
$H$}=Ih)}{\m_{R,\t,-p}(\hbox{\rm pattern of K}=Ik)}\Eq(5.1)$$
for large $\t$. If the system verifies axiom C the same result holds
with $IP$ (see \equ(4.2) replacing $I$ (without requiring any relation
between the total phase space contraction rate and the rate of
contraction of the surface of the attractor).}
\*

In other words the relative probability of fluctuation patterns of $H$
and $K$ observed in a time span $\t$ during which the average entropy
creation rate is $p\,\media{\s}_+$ are the same as those of the time
reversed patterns in a time span $\t$ in which the average entropy
creation rate is the opposite: $-p\,\media{\s}_+$.
\*

This allows us to give a physical interpretation to $p$: namely if we
look at the evolution on time laps of size $\t$ we see that the average
entropy creation rate $p$ will be usually $p=1$ and the probability of
observing $p\ne1$ will be rare and the fraction of times we shall
observe it is $e^{(\z(p)-\z(1))\t}$: hence events in which $p\ne1$ will
be rare and random (\ie {\it intermittent}) and they will take place at
rate $\z(1)-\z(p)$.
\*

The above theorem shows that when $p$ is significantly different from
$1$ ``{\it things go very wrong}''. The frequency of findings of a time
interval of size $\t$ during which the time reversed patterns are
relatively as probable as the normal patterns will be given by
$e^{-\media{\s}_+\,\t}$ {\it no matter which observable $H$ we look at}: an
independence property that can be checked, in principle, in an
experiment.

Hence a physical interpretation of $p$ is that it is a quantitative
measurement of the degree of reversibility that is observed. The larger
$1-p$ is the more ``{\it unintuitive behavior}'' will be observed.  For
$p=-1$ everything would be dramatically different from what
expected.\annota{1}{\nota ``If entropy creation rate could be changed in
sign for a minute around Niagara falls then during that minute their
water would be more likely to go up rather than down''. One ``just'' has
to change the sign of the entropy creation rate, no extra effort
needed!}  The time intervals during which anomalous behavior is
observed are rare so that their manifestation is intermittent and we
call this phenomenon ``{\it entropy driven intermittency}'': the
function $\z(p)$ describes the phenomenon quantitatively.
\*

\0{\bf\S6. Entropy driven intermittency. Observability.}
\numsec=6\numfor=1\*

We now address the question: ``is this intermittency observable''? is
its rate function $\z(p)$ measurable?

Clearly $\s_+$ and $\z(p)$ will grow with the size of the system \ie with
the number of degrees of freedom, at least, which $\tende{R\to\io}\io$
so that there should be serious doubts about the observability of so
rare fluctuations.
\*

However if we look at a small subsystem in a little volume $V_0$ of
linear size $L_0$ we can regard it again as a fluid enclosed in a box
$V_0$ described by the same reversible GNS equations. We can imagine,
therefore, that this small system also verifies a fluctuation relation
in the sense that if, \cfr \equ(4.7), \equ(3.3)

$$\eqalign{
&\s_{V_0}(\V u)= \KK_{L_0}(R)\,\n_G(\uu)\cr
&\n_G(\uu)=\fra{\ig_{V_0}\big(\V\f\cdot\D\uu-R\D\uu\cdot(\W
u\cdot\W\dpr\uu)\big)\,d\xx} {\ig_{V_0}(\D\uu)^2\,d\xx}\cr}\Eq(6.1)$$
then it should be that the fluctuations of $\s$ averaged over a time
span $\t$ are controlled by rate functions $\z_V(p)$ and $\z_{V_0}(p)$
that we can expect to be, for $R$ large

$$\eqalign{
&\z_V(p)=\lis\z(p)\,\KK_L(R), \qquad {\rm and}\cr
&\z_{V_0}(p)=
\lis\z(p)\,\KK_{L_0}(R),\cr
&\media{\s_{V_0}}_+=\lis\s_+
\,\KK_{L_0}(R)\cr}\Eq(6.2)$$
We recall that, \cfr \equ(4.7), $\KK_{L_0}(R)=\sum_{|\kk|<
K(R_{L_0})}\kk^2$ where $R_{L_0}$ is the Reynolds number on scale $L_0$
which from the OK41 theory is $R_{L_0}=(L_0/L)^{4/3} R$. So that
$\KK_{L_0}(R)\propto (L_0/L)^3 R^{15/4}$. Hence if we consider
observables dependent on what happens inside $V_0$ and if $L_0$ is small
so that $\KK_{L_0}(R)$ is not too large and we observe them in time
intervals of size $\t$ then the time frequency during which we can
observe a deviation ``of size'' $1-p$ from irreversibility will be small
of the order of

$$e^{(\lis\z(p)-\lis\z(1))\,\t\,\KK_{L_0}(R)}\Eq(6.3)$$
for $\t$ large, where the local fluctuation rate $\lis\z(p)$ verifies
(assuming transitivity or axiom C)
$$\lis\z(-p)=\lis\z(p)-\lis\s_+\,p\,\th\Eq(6.4)$$
with $\th=1$ in the transitive case and perhaps $\ne1$ when the
attracting set is smaller than phase space.
\*

Therefore by observing the frequency of intermittency one can gain
some access to the function $\lis\z(p)$.

Note that one {\it will necessarily observe a given fluctuation
somewhere} in the fluid if $L_0$ is taken small enough: in fact the
entropy driven intermittency takes place not only in time but also in
space. Thus we shall observe inside a box of size $L_0$ ``somewhere''
in the total volume $V$ of the system a fluctuation of size $1-p$ with
high probability if

$$(L/L_0)^3 e^{(\lis\z(p)-\lis\z(1))\,\t\,\KK_{L_0}(R)}\simeq 1\Eq(6.5)$$
and the special event $p=-1$ will occur with high probability if
$$(L/L_0)^3 e^{-\lis\s_+\,\t\,\KK_{L_0}(R)}\simeq 1\Eq(6.6)$$
by \equ(6.4). Once this event is realized the fluctuation patterns
will have relative probabilities as described in \S5.
\*

An attempt at interpreting the experiment performed by Ciliberto and
Laroche on convecting water at room temperature in terms of the above
theory is in [Ga99d].
\*

The idea and the possibility of local fluctuation theorems has been
developed and tested first numerically, [GP99], and then
theoretically, [Ga99c], by showing that it indeed works at least in
some models (with homogeneous dissipation like the GNS and NS
equations) which are simple enough to allow us to build a
mathematically complete theory of the phase space contraction
fluctuations.

Of course if the quantity $\media{\s_V}_+$ and $p\,\media{\s_V}_+$ could
be measurable, like the equilibrium entropy, in terms of heat ceded to
the various thermostats acting on the system divided by their
temperature, then the whole theory could be even more easily subjected
to experimental check as we could directly measure the rate function
$\z(p)$ at least in small (but still macroscopically large)
subvolumes. But the interpretation of the phase space contraction as a
``physical entropy'' (a concept that, however, we mentioned as {\it
still requiring a definition} in non equilibrium physics) is quite
controversial, \eg see [Ho99] p. 236 and p. 240, as any statement about
entropy is doomed to be.

I adhere to the point of view, [An82], [GR97], [Ru99a], that the {\it right
definition} of entropy creation rate in systems out of equilibrium is
just the phase space contraction rate: but the connection with
measurable entities of the quantity so defined is (therefore) an open
problem. Nevertheless we have seen that there are already quite a few
checks to test the theory that are already possible (and necessary
given the large number of assumptions that must be made to obtain
them).

\*
{\bf Acknowledgements:\it Work partially supported by Rutgers
University and MPI via a grant \# ?????.}

\*

{\bf References.}
\*
\def\0{\vskip3mm\noindent}

\0[An82] Andrej, L.: {\it The rate of entropy change in
non--Hamiltonian systems}, Physics Letters, {\bf 111A}, 45--46, 1982.
And {\it Ideal gas in empty space}, Nuovo Cimento, {\bf B69}, 136--144,
1982. See also {\it The relation between entropy production and
$K$--entropy}, Progress in Theoretical Physics, {\bf75}, 1258--1260,
1986.

\0[Bo66] Boltzmann, L.: {\it\"Uber die mechanische Bedeutung des
zweiten Haupsatzes der W\"armetheorie}, in
"Wissenschaftliche Abhandlungen", ed. F. Hasen\"ohrl,
vol. I, p. 9--33, reprinted by Chelsea, New York.

\0[Bo84] Boltzmann, L.: {\it \"Uber die Eigenshaften monzyklischer
und anderer damit verwandter Systeme}, in "Wissenshafltliche
Abhandlungen", ed. F.P. Hasen\"ohrl, vol. III, p. 122--152,
Chelsea, New York, 1968, (reprint).

\0[BG97] Bonetto, F., Gallavotti, G.: {\it Reversibility, coarse graining
   and the chaoticity principle}, Communications in Mathematical
   Physics, {\bf189}, 263--276, 1997.

\0[BGG97] Bonetto, F., Gallavotti, G., Garrido, P.: {\it Chaotic
   principle: an experimental test}, Physica D, {\bf 105}, 226--252,
   1997.

\0[CG99] Cohen, E.G.D., Gallavotti, G.: {\it Note on Two Theorems in
   Nonequilibrium Statistical Mechanics}, Journal of Statistical
   Physics, {\bf 96}, 1343--1349, 1999.

\0[CL98] Ciliberto, S., Laroche, C.: {\it An experimental
verification of the Galla\-vot\-ti--Cohen fluctuation theorem}, Journal de
Physique, {\bf8}, 215--222, 1998.

\0[DM96] Dettman, C.P., Morris, G.P.: {\it Proof of conjugate pairing
for an isokinetic thermostat}, Physical Review {\bf 53 E}, 5545--5549,
1996.

\0[ECM93] Evans, D.J.,Cohen, E.G.D., Morriss, G.P.: {\it Probability
of second law violations in shearing steady f\/lows}, Physical Review
Letters, {\bf 71}, 2401--2404, 1993.

\0[GC95] Gallavotti, G., Cohen, E.G.D.: {\it Dynamical ensembles in
   nonequilibrium statistical mechanics}, Physical Review Letters,
   {\bf74}, 2694--2697, 1995. And: {\it Dynamical ensembles in
   stationary states}, Journal of Statistical Physics, {\bf 80},
   931--970, 1996].

\0[Ga96a] Gallavotti, G.: {\it Chaotic hypothesis: Onsager reciprocity and 
   fluctuation--dissipa\-tion theorem},
   Journal of Statistical Phys., {\bf 84}, 899--926, 1996.

\0[Ga96b] Gallavotti, G.: {\it New methods in nonequilibrium gases and
fluids}, Open Systems and Information Dynamics, Vol. 6, 101--136, 1999
(original in chao-dyn \#9610018).

\0[Ga97a] Gallavotti, G.: {\it Dynamical ensembles equivalence in fluid
   mechanics}, Physica D, {\bf 105}, 163--184, 1997.

\0[Ga97b] Gallavotti, G.: {\it Ipotesi per una introduzione alla Meccanica
  dei Fluidi}, ``Quader\-ni del CNR-GNFM'', vol. {\bf 52}, p. 1--428,
  Firenze, 1997. English translation in progress: available at
  http:$\backslash\backslash$ipparco.roma1.infn.it.

\0[Ga98a] Gallavotti, G.: {\it Chaotic dynamics, fluctuations,
   non-equilibrium ensembles},\\ Chaos, {\bf8}, 384--392, 1998. See also
   [Ga96c].

\0[Ga98b] Gallavotti, G.: {\it Breakdown and regeneration of time reversal
  symmetry in non\-equi\-librium Statistical Mechanics}, Physica D, {\bf112},
  250--257, 1998.

\0[Ga99a] Gallavotti, G.: {\it Statistical Mechanics}, Springer Verlag,
1999.

\0[Ga99b] Gallavotti, G.: {\it Fluctuation patterns and conditional
  reversibility in nonequilibrium systems}, Annales de l' Institut
  H. Poincar\'e, {\bf70}, 429--443, 1999.

\0[Ga99c] Gallavotti, G.: {\it A local fluctuation theorem}, Physica A,
{\bf 263}, 39--50, 1999. And {\it Chaotic Hypothesis and Universal Large
Deviations Properties}, Documenta Mathematica, extra volume ICM98,
vol. I, p. 205--233, 1998, also in chao-dyn 9808004.

\0[Ga99d] Gallavotti, G.: {\it Ergodic and chaotic hypotheses:
nonequilibrium ensembles in statistical mechanics and turbulence},
chao-dyn \# 9905026; and {\it Non equilibrium in statistical and fluid
mechanics.  Ensembles and their equivalence.  Entropy driven
intermittency.}, chao-dyn \# 0001???.

\0[Ge98] Gentile, G.: {\it Large deviation rule for Anosov flows},
Forum Mathematicum, {\bf10}, 89--118, 1998.

\0[GP99] Gallavotti, G., Perroni, F.: {\it An experimental test of the
local fluctuation theorem in chains of weakly interacting Anosov
systems}, preprint, 1999, in http://ipparco. roma1. infn. it at the 1999
page.

\0[GR97] Gallavotti, G., Ruelle, D.: {\it SRB states and
non-equilibrium statistical mechanics close to equilibrium},
Communications in Mathematical Physics, {\bf190}, 279--285, 1997.

\0[Ho99] Hoover, W. G.: {\it Time reversibility, Computer simulation,
and Chaos}, World Scientific, 1999.

\0[LL71] Landau, L., Lifchitz, E.: {\sl M\'ecanique des fluides}, MIR,
Moscou, 1971.

\0[RS99] Rondoni, L., Segre, E.: {\it Fluctuations in two dimensional
reversibly damped turbulence}, Nonlinearity, {\bf12}, 1471--1487, 1999.

\0[Ru76] Ruelle, D.: {\it A measure associated with Axiom A
attractors}, American Journal of Mathematics, {\bf98}, 619--654, 1976.

\0[Ru78a] Ruelle, D.: {\it Sensitive dependence on initial conditions
and turbulent behavior of dynamical systems}, Annals of the New York
Academy of Sciences, {\bf356}, 408--416, 1978. This is the first place
where the hypothesis analogous to the later chaotic hypothesis was
formulated (for fluids): however the idea was exposed orally at least
since the talks given to illustrate the technical work [Ru76], which
appeared as a preprint and was submitted for publication in 1973 but was
in print three years later.

\0[Ru96] Ruelle, D.: {\it Positivity of entropy production in
nonequilibrium statistical mechanics}, Journal of Statistical Physics,
{\bf 85}, 1--25, 1996. And Ruelle, D.: {\it Entropy production in
nonequilibrium statistical mechanics}, Communications in Mathematical
Physics, {\bf189}, 365--371, 1997.

\0[Ru99a] Ruelle, D.: {\it Smooth dynamics and 
new theoretical ideas in non-equilibrium statistical mechanics},
Journal of Statistical Physics, {\bf 95}, 393--468, 1999.

\0[Ru99b] Ruelle, D.: {\it A remark on the equivalence of isokinetic and
isoenergetic thermostats in the thermodynamic limit}, preprint, to
appear in Journal of Statistical Physics, 1999.

\0[WL98] Woitkowsky, M.P., Liverani, C.: {\it Conformally Symplectic
Dynamics and Symmetry of the Lyapunov Spectrum}, Communications in
Mathematical Physics, {\bf 194}, 47--60, 1998.

\def\FINE{
\*
\*
\0{\it Internet:
Author's  preprints at:  {\tt http://ipparco.roma1.infn.it}

\0\sl e-mail: {\tt giovanni.gallavotti@roma1.infn.it}
}}
\FINE

\end